\documentclass[letterpaper, 10 pt, conference]{ieeeconf}  

\IEEEoverridecommandlockouts                              

\usepackage{amsmath} 
\usepackage[utf8]{inputenc}

\usepackage{graphicx}
\usepackage[version=4]{mhchem}
\usepackage{siunitx}
\usepackage{longtable,tabularx}
\usepackage{amsfonts,color,soul}
\usepackage{textcomp, comment, algorithm, algpseudocode}
\usepackage{xcolor, bbm}
\usepackage{hyperref}
\usepackage{multirow}
\makeatletter
\newcommand\fs@betterruled{%
  \def\@fs@cfont{\bfseries}\let\@fs@capt\floatc@ruled
  \def\@fs@pre{\vspace*{7pt}\hrule height.8pt depth0pt \kern2pt}%
  \def\@fs@post{\kern2pt\hrule\relax}%
  \def\@fs@mid{\kern2pt\hrule\kern2pt}%
  \let\@fs@iftopcapt\iftrue}
\floatstyle{betterruled}
\restylefloat{algorithm}
\makeatother

\title{\LARGE \bf
Path Integral Control with Rollout Clustering and Dynamic Obstacles
}

\author{Steven Patrick$^{1}$ and Efstathios Bakolas$^{1}$
\thanks{*This research has been supported in part by NSF award ECCS-1924790.}
\thanks{$^{1}$The University of Texas at Austin, Austin, Texas 78712-1221 
    {\tt\small spatric5@utexas.edu},  and {\tt\small bakolas@austin.utexas.edu}}%
}

\begin{document}

\maketitle
\thispagestyle{empty}
\pagestyle{empty}

\begin{abstract}
Model Predictive Path Integral (MPPI) control has proven to be a powerful tool for the control of uncertain systems (such as systems subject to disturbances and systems with unmodeled dynamics). One important limitation of the baseline MPPI algorithm is that it does not utilize simulated trajectories to their fullest extent. 
For one, it assumes that the average of all trajectories weighted by their performance index will be a safe trajectory.
In this paper, multiple examples are shown where the previous assumption does not hold, and a trajectory clustering technique is presented that reduces the chances of the weighted average crossing in an unsafe region.
Secondly, MPPI does not account for dynamic obstacles, so the authors put forward a novel cost function that accounts for dynamic obstacles without adding significant computation time to the overall algorithm.
The novel contributions proposed in this paper were evaluated with extensive simulations to demonstrate improvements upon the state-of-the-art MPPI techniques.
\end{abstract}



\section{Introduction}\label{sec:Intro}
For autonomous agents to be useful in unstructured environments, motion planning algorithms \cite{lavalle2006planning} must be used to ensure avoidance of obstacles (both static and dynamic) at all times. Additionally, the algorithms must be robust to a variety of different disturbances introduced by the real world: actuation noise, measurement error, process noise, and unmodeled dynamics.

A class of trajectory optimization algorithms that has been developed to address a subset of these sources of uncertainty is Model Predictive Path Integral (MPPI) control \cite{williams2016aggressive}. The key idea for MPPI control is to simulate multiple sequences of control inputs over a given time horizon and initial condition to determine an optimal control sequence.
With each simulation, a realization from a random variable is used to perturb the original input sequence.
The resulting trajectory is evaluated with a cost function, and then all of the results from the independent simulations are combined to create a control input sequence robust to the real-world disturbances.
The benefits of this algorithm are the speed of planning, robustness to control noise, and versatility to different agent dynamics \cite{williams2016aggressive}.
MPPI has been applied to several robotics problems, including racing on a dirt track \cite{williams2017Information}, map-less navigation \cite{Mohamed2022Autonomous}, drone flights \cite{mohamed2020model}, and manipulation \cite{bhardwaj2022storm} to name a few.

\textit{Literature Review}: Several improvements upon the baseline of MPPI have been proposed.
One category of improvements change how the samples are generated.
Instead of using a handcrafted static distribution to sample from, other works have used non-static distributions \cite{balci2022constrained} or learned distributions \cite{sackslearning} to generate trajectories.
Another improvement is narrowing the search space of control inputs using control barrier functions \cite{tao2022path}.
A major improvement from the original MPPI is accounting for noise that does not enter the system dynamics through the control input channels. In \cite{williams2018robust}, the authors account for process noise by having a nominal system and a real system. 
This allows for a larger exploration space of the MPPI algorithm, and it highlights how sampling from distributions other than the standard Gaussian produces better results.
In \cite{yin2022risk}, process noise was added to the simulated trajectories and the likelihood of the state disturbance was directly accounted for in the cost function.
Even though their system was shown to be more robust than other MPPI variants, the number of trajectories to be simulated was significantly greater.
The increase in simulated trajectories means the algorithm can only be applied to systems with a GPU to plan in real-time.

\begin{figure}
    \centering
    \includegraphics[scale=.22]{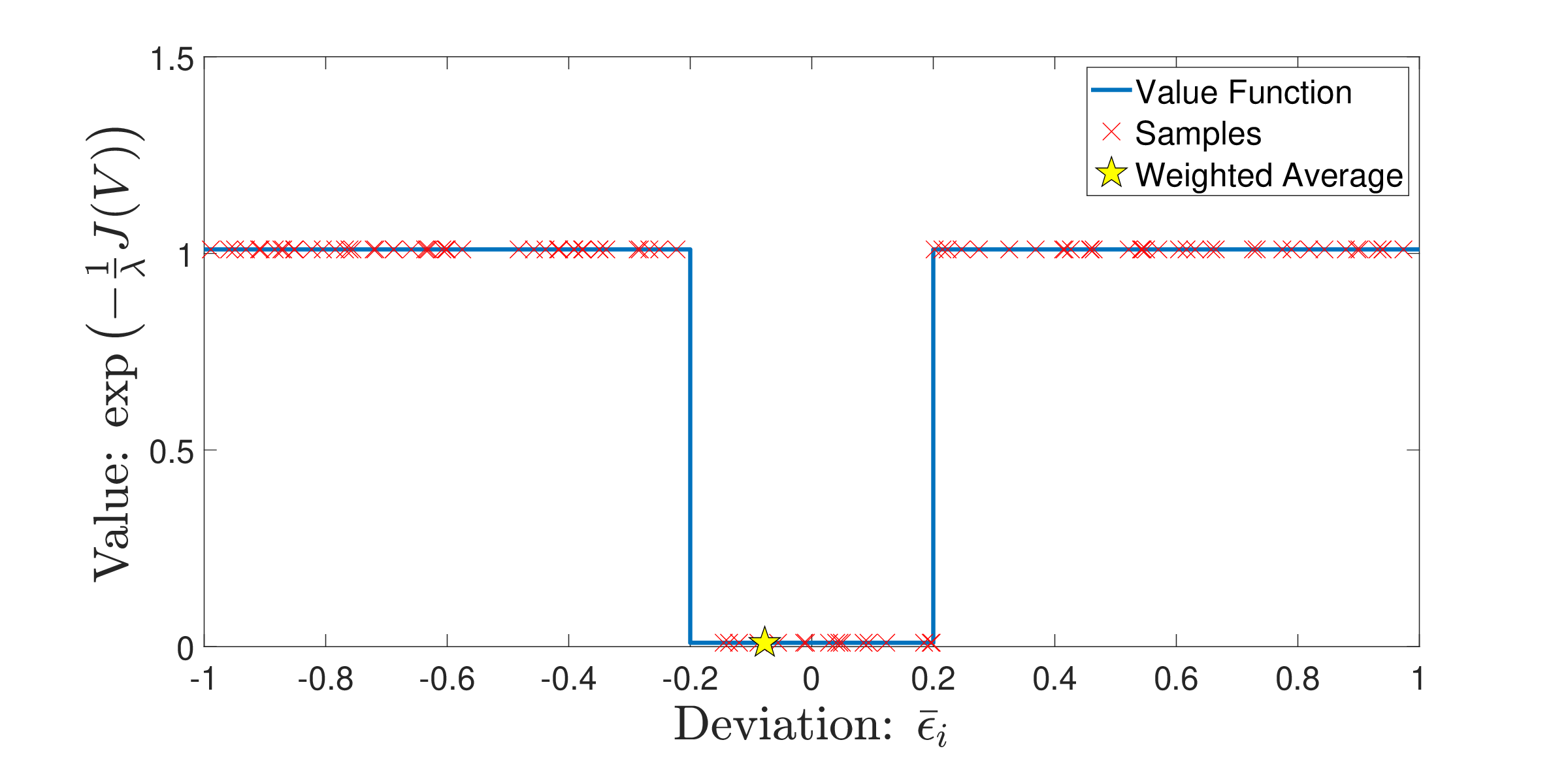}
    \caption{Example of standard MPPI failing. Trajectories with low cost, and therefore high value, are separated by a region with high cost trajectory. The resulting weighted average used by MPPI is in the high cost, low value, region.}
    \label{fig:MPPI Failure}
\end{figure}

Even with these improvements, there are still cases where MPPI performs poorly. 
Figure \ref{fig:MPPI Failure} illustrates one such problem where the standard weighted average scheme results in a control input that is in a region of low value for the associated cost function.
Another disadvantage of using the baseline MPPI method is it was developed for static environments. 
When applied to dynamic environments, quick replanning is used to account for moving obstacles.
However, this type of planning is likely to fail if the obstacles move quickly or if the agent is close to the moving obstacle.
Both cases are likely to result in a collision since the agent is planning under the false assumption that the world is static.

\textit{Contributions}: To account for both of these issues, two novel methods are proposed to augment MPPI (in fact, the proposed improvements can be applied to any state-of-the-art variant of MPPI). The first is based on clustering trajectories and using MPPI only on clusters of trajectories.
The second is sampling trajectories from moving obstacles and incorporating their probability of occurring into the cost of the trajectory.
The approach in this paper would be similar to \cite{yin2022risk} if they augmented the state of the system with dynamic obstacles but with a key difference.
In this paper, the cost function is parameterized by simulated obstacle trajectories which reduces the computational load to account for dynamic obstacles.

\textit{Outline}: In Section \ref{sec:Background}, a brief review of the baseline MPPI algorithm is made. After that, the proposed improvements upon the baseline MPPI algorithm in Section \ref{sec:Methodology} are denoted. Then, the improvements are demonstrated in multiple environments through non-trivial simulations in Section \ref{sec:Results}. Finally, in Section \ref{sec:Conclusion} the results are discussed and future improvements upon the proposed algorithm are expressed.

\section{Background}\label{sec:Background}
\subsection{Model Predictive Path Integral  (MPPI)}
Consider an agent whose equations of motion can be described by the following discrete-time state space model:
\begin{gather} \label{eq:Difference Equation}
    \mathbf{x}_{i+1} = f(\mathbf{x}_i,\mathbf{u}_i),
\end{gather}
where $\mathbf{x}_i \in \mathbb{R}^n$ is the current state at time step $i$ and $\mathbf{u}_i \in \mathbb{R}^m$ is the control input to be applied at the same time step.
MPPI uses Eq. (\ref{eq:Difference Equation}) to simulate several possible trajectories given a sequence of control inputs over a time horizon, $N$. 
The variable $U$ is used to denote a sequence of control inputs $U = \{\mathbf{u}_{0},\mathbf{u}_1, \dots, \mathbf{u}_{N-1} \}$ whose  application results in a state sequence, $X = \{\mathbf{x}_0,\dots,\mathbf{x}_{N}\}$, for a given initial condition, $\mathbf{x}_0$. 
In addition, $\tau = (X,U)$ denotes the combined state and input sequences and is referred to as an agent's trajectory. 

In the real world, Eq. (\ref{eq:Difference Equation}) is an inaccurate model due to assuming the control inputs are applied without noise.
MPPI accounts for this by assuming the disturbance to desired control input is effected by additive noise: $\mathbf{v}_i=\mathbf{u}_i+\boldsymbol{\epsilon}_i$ 
where $\boldsymbol{\epsilon}_i$ is a random normal variable with positive definite variance, $\Sigma$, and zero mean.
The realization of this control input is determined by the distribution, $\mathbb{P}$, referred to as the uncontrolled distribution, whose probability density function (PDF) is given by
\begin{gather} \label{eq:uncontrolled distribution}
    p(V) = \prod_{i=0}^{N-1} ((2\pi)^m |\Sigma|)^{-1/2} \exp \left ( -\frac{1}{2} \mathbf{v}_i^T \Sigma^{-1} \mathbf{v}_i \right ) \\
    \text{where }V = \{\mathbf{v}_0, \mathbf{v}_1, \cdots, \mathbf{v}_{N-1}\}.
\end{gather}
As seen in Eq. (\ref{eq:uncontrolled distribution}), the mean is zero because no control input has been applied to the system, which is the reason why the distribution $\mathbb{P}$ is referred to as the uncontrolled distribution. By contrast, the controlled distribution, which is denoted as $\mathbb{Q}$, has the noiseless control input as its mean.
The PDF of $\mathbb{Q}$ is therefore given by
\begin{align} \label{eq:controlled distribution}
    q(V) = \prod_{i=0}^{N-1} &((2\pi)^m |\Sigma|)^{-1/2}  \notag\\
    & \exp \left ( -\frac{1}{2}(\mathbf{v}_i-\mathbf{u}_i)^T \Sigma^{-1} (\mathbf{v}_i-\mathbf{u}_i) \right ).
\end{align}
To determine the optimal control input given these potential disturbances, MPPI minimizes the expectation of a cost function over $\mathbb{P}$. In particular,
\begin{align} \label{eq:stochastic min}
    U^* &= \arg \min_{U} \mathbb{E}_{\mathbb{P}} \left ( J(V) \right ) \\ 
    J(V) &= \phi(\mathbf{x}_N) + \sum_{i=0}^{N-1} \psi(\mathbf{x}_i,\mathbf{v}_i) \notag\\
    \label{eq:Cost Function}
    &= \phi(\mathbf{x}_N) + \sum_{i=0}^{N-1} \psi(\mathbf{x}_i,\mathbf{u}_i+\boldsymbol{\epsilon}_i)
\end{align}
where $\phi$ is the terminal cost function, and $\psi$ is the running cost function.
In \cite{theodorou2012relative}, it was shown that solving the stochastic optimal control problem given in (\ref{eq:stochastic min}) is equivalent to finding the minimum KL-Divergence between the controlled and uncontrolled distributions.
The relationship was used to derive a new distribution, $\mathbb{Q}^*$, for importance sampling defined by the following PDF:
\begin{gather}\label{eq:Optimal Density Function}
    q^*(V) = \frac{1}{\eta} \exp \left ( -\frac{1}{\lambda} J(V)\right ) p(V),
\end{gather}
where $\eta>0$ is a normalizing constant such that the integral of the PDF, $q^*(V)$, over the sample space is $1$. 
The sensitivity parameter, $\lambda>0$, allows the user to set how important differences in cost between two trajectories are. 
A large $\lambda$ results in low sensitivity to cost differences and small $\lambda$ produce high sensitivity.

This new distribution allows for the control input to be defined as
\begin{gather} \label{eq:optimal integral}
    \mathbf{u}_i = \int q^*(V) \mathbf{v}_i \mathrm{d}V
\end{gather}
With Eq. (\ref{eq:optimal integral}), importance sampling is used to optimize the number of samples needed to get a reliable approximation of the optimal control input.
To accomplish this, Eq. (\ref{eq:optimal integral}) is multiplied by PDFs that are strictly positive whenever $q^*(V)\mathbf{v}_i\neq 0$. 
Clearly, Eq. (\ref{eq:uncontrolled distribution}) and (\ref{eq:controlled distribution}) both satisfy this condition and using them for importance sampling results in the following
\begin{align} \label{eq:new integral}
    \mathbf{u}^*_i &= \int \frac{q^*(V)}{p(V)}\frac{p(V)}{q(V)} q(V) \mathbf{v}_i \mathrm{d}V.
\end{align}
The next step is to define the so-called weighting function, which is denoted as $w(V)$ and defined as
\begin{align} \label{eq:weight fraction function}
    w(V) = \frac{q^*(V)}{p(V)}\frac{p(V)}{q(V)}.
\end{align}
Substituting in the weighting function into Eq. (\ref{eq:new integral}) results in the new optimal control input calculated by
\begin{gather} \label{eq: expected value}
    \mathbf{u}^*_i = \mathbb{E}_{\mathbb{Q}} [w(V)\mathbf{v}_i].
\end{gather}
The ratios $q^*(V)/p(V)$ and $p(V)/q(V)$ used to define $w(V)$ in Eq. (\ref{eq:weight fraction function}) are known and given by
\begin{align} \label{eq:optimal over natural weighting}
    \frac{q^*(V)}{p(V)} &= \frac{1}{\eta} \exp \left ( -\frac{1}{\lambda} J(V)\right ) \\ 
    \label{eq:natural over good weighting}
    \frac{p(V)}{q(V)} &= \exp \left (\sum_{i=0}^{N-1} \frac{1}{2}\mathbf{u}_i^T \Sigma^{-1} \mathbf{u}_i- \mathbf{v}_i^T \Sigma^{-1} \mathbf{v}_i \right ).
\end{align}
Substituting Eq. (\ref{eq:optimal over natural weighting}) and (\ref{eq:natural over good weighting}) into Eq. (\ref{eq:weight fraction function}) results in
\begin{align} \label{eq: OG weighting func}
    Q(V) &= \sum_{i=0}^{N-1} \frac{1}{2}\mathbf{u}^{(k)T}_i \Sigma^{-1} \mathbf{u}_i^{(k)} 
     - \mathbf{v}_i^{(k)T} \Sigma^{-1} \mathbf{v}^{(k)}_i \\ \label{eq:first weight equation}
    w(V) &= \frac{1}{\eta} \exp \left ( -\frac{1}{\lambda} J(V) + Q(V) \right ) .
\end{align}
In Eq. (\ref{eq:first weight equation}), $\eta$ is the only remaining unknown.
However, it is is computationally intractable to determine the normalizing constant for the PDF in Eq. (\ref{eq:Optimal Density Function}).
Instead, a Monte-Carlo approach \cite{tao2022path} is used wherein $K$ realizations of control disturbances are sampled.
\begin{align} \label{eq:first norm equation}
    \eta & \approx \sum_{k=1}^{K} \exp \left ( -\frac{1}{\lambda} J(V_{k})  + Q(V_k)  \right ).
\end{align}
\begin{figure*}
    \centering
    \includegraphics[width=.9\textwidth]{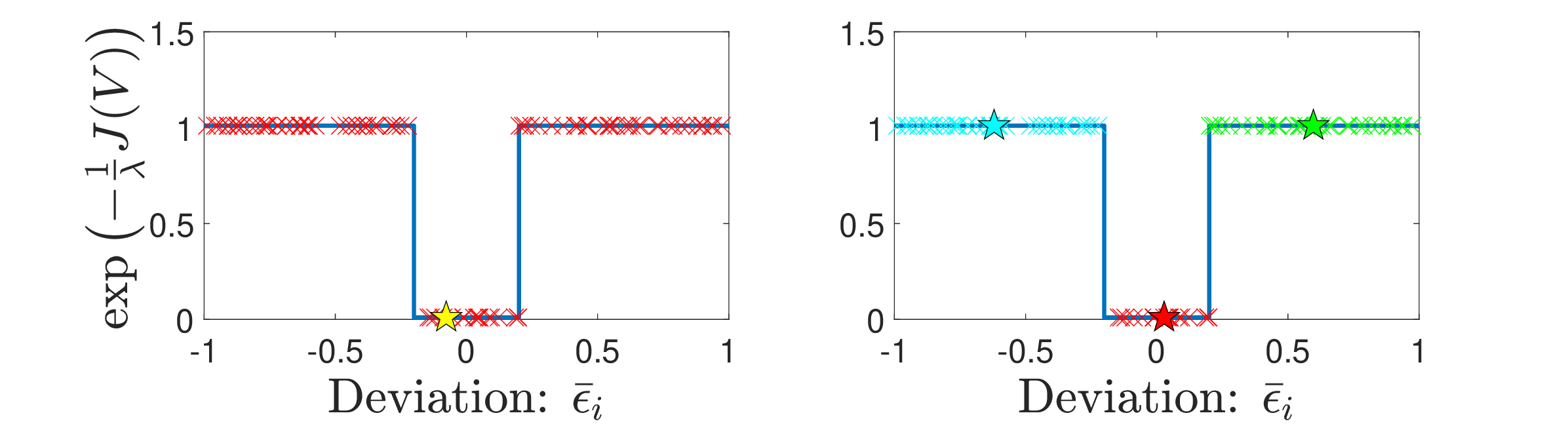}
    \caption{Example of standard MPPI (left) producing an undesirable result and Clustered MPPI (right) producing multiple valid solutions. The blue solid line is the cost function, the X marks are sampled points, and the filled in stars are the result of the weighted average.}
    \label{fig:Cluster Example}
\end{figure*}
The $k$-th sample, $\mathcal{E}_k$, is used to generate a sequence of noisy control inputs $V_k$ from a noiseless control reference $U_k$:
\begin{align}
    \mathcal{E}_k &= \{\boldsymbol{\epsilon}^{(k)}_{0},\cdots,\boldsymbol{\epsilon}^{(k)}_{N-1}\} \\
    U_{k} &= \{\mathbf{u}^{(k)}_{0},\cdots,\mathbf{u}^{(k)}_{N-1}\} \\
    V_k &= U_k+\mathcal{E}_k.
\end{align}
In theory, $U_k$, can be sampled arbitrarily from the allowed bounds of the control inputs.
In practice, it is more efficient to use a single reference control input $U_{0}$ and weight the perturbations with the following equations
\begin{align}
    \label{eq:weight update}
    R(\mathcal{E}_k) &= \sum_{i=0}^N \frac{1}{2} \mathbf{u}_{i}^{(0)T} \Sigma^{-1} (\mathbf{u}_{i}^{(0)}+2 \boldsymbol{\epsilon}^{(k)}_{i}) \\
    w(\mathcal{E}_k) &= \frac{1}{\eta} \exp \left ( -\frac{1}{\lambda}J(U_0+\mathcal{E}_k) - R(\mathcal{E}_k) \right ) \\ \label{eq:second norm equation}
    \eta &= \sum_{k=1}^K \exp \left ( -\frac{1}{\lambda}J(U_0+\mathcal{E}_k)  - R(\mathcal{E}_k) \right ) \\
    \label{eq:control update}
    u^*_{i} &= \mathbf{u}^{(0)}_{i} + \sum_{k=1}^K w(\mathcal{E}_k) \boldsymbol{\epsilon}^{(k)}_i
\end{align}
It can be seen that Eq. (\ref{eq: OG weighting func})-(\ref{eq:first norm equation}) is equivalent to Eq. (\ref{eq:weight update})-(\ref{eq:second norm equation} when all $U_k$ are equal.
The use of Eq. (\ref{eq:control update}) to determine the final control input sequence can be seen in Algorithm \ref{alg:standard MPPI}.
In this algorithm, there is an extra variable $\rho$ not seen in Eq. (\ref{eq:weight update})-(\ref{eq:control update}).
The purpose of this variable is to render the algorithm more stable by preventing rounding errors in calculating the exponent of large negative numbers \cite{williams2018robust}.
\begin{algorithm}
\caption{MPPI Control Algorithm} \label{alg:standard MPPI}
\begin{algorithmic}[1]
\Require{$U_0, \{\mathcal{E}_1,\cdots, \mathcal{E}_K\}, \{S_1, \cdots , S_K\}, \lambda$} \Comment{Reference Input, Perturbations, Cost of Trajectories, Sensitivity}
\State {$\rho \gets \min(S_1, \cdots , S_k)$}
\State{$\eta \gets \sum_{k=1}^K \exp\left (-\frac{1}{\lambda} (S_k-\rho) \right )$}
\For {$k \gets 1 \text{ to } K$}
    \State{$w_k = \frac{1}{\eta}\exp\left (-\frac{1}{\lambda} (S_k-\rho) \right )$}
\EndFor
\State{$U \gets U_0 + \sum_{k=1}^K w_k \mathcal{E}_k$}
\end{algorithmic}
\end{algorithm}

\section{Methodology}\label{sec:Methodology}

\subsection{Rollout Clustering}
As mentioned in Section \ref{sec:Intro}, MPPI can perform poorly when the value function has multiple local maxima. 
To address this limitation of MPPI, clustering rollouts into groups that do not contain sharp discontinuities in the cost function is proposed.
The authors' chosen clustering algorithm is a density-based method \cite{campello2020density} called DBSCAN \cite{ester1996density}.
The algorithm is a natural choice for this problem because perturbations that are close together and produce similar costs will be clustered together, but sharp changes in the cost will cause a new cluster boundary to form.
Clustering in this manner prevents undesirable valleys separating high-points of the value function from affecting the expected value computation.
Since the agent does not know the number of valleys or high points along the value function, this method was also chosen because it allows for a dynamic number of clusters unlike K-means \cite{hartigan1979algorithm}.

DBSCAN works by sequentially building clusters as the data comes in.
For a new data point, all previous clusters are checked against a ball of radius $\epsilon$ centered at the new data point.
If there are any intersections, the point is either added to an existing cluster, or multiple clusters are merged that intersect with the ball.
If no intersections are formed, then the data-point is its own cluster.
This process is repeated until all the data points have been processed.
The data point, $\mathbf{d}_k$, for the $k$-th rollout are the perturbations from the control input, $\mathcal{E}$, and their associated trajectory cost, $J(V)$, from Eq. (\ref{eq:Cost Function}): $\mathbf{d}_k = \{\boldsymbol{\epsilon}_{N-1}^{(k)},\cdots, \boldsymbol{\epsilon}_0^{(k)}, J(V_k)\}$.
With the clustered data-points, a new distribution can be chosen to derive the importance sampling weights.
This distribution should have a high probability for trajectories within a cluster but a low probability for trajectories outside of the cluster.
However, this distribution has to satisfy two properties to be a valid distribution for importance sampling.
The first being it has to have a valid probability density function.
The second is that the probability density function must be strictly positive where the original PDF is non-zero.

To satisfy the two conditions and get the desired attributes, a truncated Gaussian is used as a distribution for importance sampling.
The Gaussian from Eq. (\ref{eq:Optimal Density Function}) is truncated such that all noise realizations within a cluster evaluate to non-zero values much greater than realizations outside of the cluster.
The resulting PDF is defined as
\begin{gather} \label{eq:Truncated Gaussian PDF}
    y(V) = \begin{cases}
    \upsilon  \exp \left ( -\frac{1}{\lambda} J(V)\right ) p(V), \quad V \in \Omega \\
    \sigma  \exp \left ( -\frac{1}{\lambda} J(V)\right ) p(V), \quad V \in \Omega^c
    \end{cases}
\end{gather}
where $\Omega$ is a simply connected compact region that encompasses all of the samples of a given cluster, and $\Omega^c$ is its complement.
Additionally, $\upsilon \gg \sigma > 0$ are normalizing constants for $y(V)$ to be a valid probability density function.
\subsection*{Proposition 1}
The truncated normal distribution, $y(V)$, is a valid importance sampling distribution.

\begin{proof} Found in Appendix. \end{proof}
\subsubsection*{Remark 1}
It can be seen that the weighting function for the truncated Gaussian distribution can be calculated using the weights from the original Gaussian distribution.
Therefore, no additional evaluations of the cost function or extra simulations are required to find the importance sampling weights of a subset of trajectories.
It is therefore unnecessary to calculate these parameters as well.
Note that the clustered MPPI approach adds little computational burden to the baseline MPPI algorithm.
\subsubsection*{Remark 2}
Just like in the original MPPI, the normalizing constant, $\upsilon$ must be estimated with Monte-Carlo methods similar to Eq. (\ref{eq:first norm equation}).
This is done using the following equation
\begin{align}
    \eta &\approx \sum_{V_k \in \Omega}  \exp \left ( -\frac{1}{\lambda} J(V_{k})  + Q(V_k)  \right )
\end{align}
The other normalizing constant, $\sigma$, is not estimated since it is chosen to be arbitrarily close to $0$.
\begin{algorithm}
\caption{Clustered MPPI}
\label{alg:Clustered MPPI}
\begin{algorithmic}[1]
\Require{$f,\Sigma$} \Comment{Difference Equation, Noise Covariance}
\Require{$\mathbf{x}_0,U_0$}\Comment{Initial Condition, Control Reference}
\Require{$K,N$}\Comment{\# of Samples, \# of Time Steps}
\Require{$\lambda$}\Comment{Cost Sensitivity}
\Require{$\psi,\phi$} \Comment{Running Cost, Terminal Cost}
\For {$k \gets 1 \text{ to } K$} \label{alg_lin:Sample For Loop} \Comment{In Parallel}
    \State $\Tilde{\mathbf{x}}^{(k)}_0\gets \mathbf{x}_0$
    \For {$i \gets 0 \text{ to } N-1$} \label{alg_lin:Traj For Loop}
        \State{$\boldsymbol{\epsilon}^{(k)}_i \gets \text{ sample } \mathcal{N}(\mathbf{0},\Sigma)$}
        \State{$\Tilde{\mathbf{x}}^{(k)}_{i+1} \gets f(\Tilde{\mathbf{x}}_i,\mathbf{u}_i+\boldsymbol{\epsilon}^{(k)}_i)$}
        \State{$S_k \mathrel{+}= \psi(\Tilde{\mathbf{x}}^{(k)}_i)+\lambda \mathbf{u}_i^T \Sigma^{-1} \boldsymbol{\epsilon}^{(k)}_i $}
    \EndFor
    \State{$S_k \mathrel{+}= \phi(\Tilde{\mathbf{x}}^{(k)}_N)$}
    \State{$X_k,\mathcal{E}_k \gets \{\Tilde{\mathbf{x}}^{(k)}_0,\cdots,\Tilde{\mathbf{x}}^{(k)}_N\},\{\boldsymbol{\epsilon}^{(k)}_0,\cdots,\boldsymbol{\epsilon}^{(k)}_{N-1}\}$}
\EndFor \label{alg_lin:End Traj For Loop}
\State{$C_1,\cdots,C_M \gets \text{ dbscan}(\{\mathcal{E}_0,S_0\}, \cdots,\{\mathcal{E}_K,S_K\})$} \label{alg_lin:dbscan}
\For {$m \gets 1 \text{ to } M$}
    \State{$U^{(m)} \gets \text{ MPPI} (U_0,\mathcal{E}_{C(m)},S_{C(m)},\lambda)$} \Comment{Alg. \ref{alg:standard MPPI}}
\EndFor \label{alg_lin:clustered_MPPI}
\State{$U \gets \arg\min_{U^{(m)}} J(U^{(m)})$} \Comment{Eq. (\ref{eq:Cost Function})}\label{alg_lin:c_MPPI selection}
\end{algorithmic}
\end{algorithm}

The pseudo-code for Clustered MPPI is given in Algorithm \ref{alg:Clustered MPPI}. Starting in Line \ref{alg_lin:Sample For Loop}, the \textbf{for} loop simulates trajectories by sampling perturbations to the reference control input and then storing the cost, trajectory, and associated noise. 
After the \textbf{for} loop in Line \ref{alg_lin:dbscan}, the trajectory cost is concatenated with its associated noise vector and the DBSCAN algorithm is performed over the $K$ data points. 
The result is an index set, $C$, defining $M$ clusters from the original samples.
The \textbf{for} loop in Line \ref{alg_lin:clustered_MPPI} calculates the new control input for each cluster using Algorithm \ref{alg:standard MPPI}.
Finally, in Line \ref{alg_lin:c_MPPI selection}, the cluster which produces the minimum cost using Eq. (\ref{eq:Cost Function}) is selected as the control input.

An illustrative example of the difference in output between the standard MPPI algorithm versus a clustered approach can be seen in Figure \ref{fig:Cluster Example}.
The $y$-axis is the value, also known as the negative exponential of the cost $J(V)$, and the $x$-axis is a deviation from a reference input.
Troughs in the value function can be encountered when the reference trajectory has the agent going straight through an obstacle.
In this example, standard MPPI fails because the weighted average between two high areas lies within the valley.
In contrast, Clustered MPPI separates these two high areas and the low area into separate clusters and performs the weighted average only including the points in the cluster.
This produces two valid selections of control input and one invalid.

\subsection{Dynamic Obstacles}
To account for dynamic obstacles, terminal and running cost functions that are parameterized by realizations of obstacle trajectories are proposed.
This is in contrast to methods like \cite{yin2022risk} where each control input deviation has an independent sampling of obstacle trajectories.
This leads to a multiplicative increase in computation time for $N$ perturbations to the control input and $P$ realizations of obstacle trajectories: $O(PN)$. 
The proposed approach is additive where the $P$ realizations are only done once which results in an additive increase in computation time: $O(P+N)$. 

To reiterate, the $l$-th obstacle has $P$ simulated trajectories collected in the set $\mathcal{O}_l$. The $p$-th trajectory of the $l$-th obstacle, $\tau_p^l$, consists of the obstacles state over the time horizon.
\begin{align}
    \tau^l_p &= \{\mathbf{o}_0^{l,p},\cdots,\mathbf{o}_N^{l,p}\} \\
    \mathcal{O}_l &= \{\tau^l_1,\cdots,\tau^l_P\}
\end{align}
The collection of $L$ dynamic obstacles trajectories, $\mathcal{O}$, is used to create a subset of the configuration space at each time step where the controlled agent is considered in collision.
\begin{align}
    \mathcal{O} &= \{\mathcal{O}_1,\cdots,\mathcal{O}_L\}
\end{align}
If a collision occurs between the controlled agent and the $p$-th trajectory of the $l$-the obstacle, then the binary function $\mathbf{1}_p^l$ equals $1$ and $0$ otherwise.
The associated trajectory also has a function, $\theta$, that outputs the probability of the trajectory occurring.
For example, this probability can be derived from Extended Kalman Filter estimate of the obstacle's position and heading of a differential drive robot.
The new running and terminal cost functions are defined as
\begin{gather} \label{eq:augmeted running cost}
    \psi(\mathbf{x}_i | \mathcal{O}) := \psi(\mathbf{x}_i) + \beta \sum_{l=1}^L \sum_{p=1}^P \theta(\tau_p^l) \mathbf{1}_p^l(\mathbf{x}_i,t_i) \\
    \label{eq:augmeted terminal cost}
    \phi(\mathbf{x}_i | \mathcal{O}) := \phi(\mathbf{x}_i)+ \beta \sum_{l=1}^L \sum_{p=1}^P \theta(\tau_p^l) \mathbf{1}_p^l(\mathbf{x}_i,t_i)
\end{gather}

In Eq. (\ref{eq:augmeted running cost}) and (\ref{eq:augmeted terminal cost}), $\beta \in \mathbb{R}^+$ is a scalar weight associated with hitting an obstacle.
With this formulation, the cost function $J(V)$ only depends on the agent's state after the obstacles' trajectories have been sampled.
This has the aforementioned benefit of reducing the computational cost of accounting for dynamic obstacles over other state-of-the-art methods.
The combination of clustered MPPI with simulating dynamic obstacles can be seen in Algorithm \ref{alg:DC-MPPI}.

\begin{algorithm}
\caption{DC-MPPI}
\label{alg:DC-MPPI}
\begin{algorithmic}[1]
\Require{$f,\Sigma,\mathbf{x}_0,U_0,K,N,\lambda$} \Comment{Alg. \ref{alg:Clustered MPPI} Parameters}
\Require{$\{\mathbf{o}_0^{(1)},\cdots,\mathbf{o}_0^{(L)}\}$} \Comment{obstacles initial states}
\Require{$\{U_o^{(1)},\cdots,U_o^{(L)}\}$} \Comment{obstacle reference inputs}
\Require{$g,\Sigma_o$} \Comment{obstacle difference equation, obstacle input variance}
\For {$l \gets 1 \text{ to } L$} \label{alg_lin:obs simulation}
    \For {$p \gets 1 \text{ to } P$}
        \State{$\{\mathbf{u}_0^{l,p},\cdots,\mathbf{u}_{N-1}^{l,p}\} \gets U_o^{(l)}$}
        \State{$\mathbf{o}_0^{l,p} \gets \mathbf{o}_0^{(l)}$}
        \For {$i \gets 0 \text{ to } N-1$}
            \State{$\boldsymbol{\epsilon} \gets \text{ sample } \mathcal{N}(0,\Sigma_o)$}
            \State {$\mathbf{o}_{i+1}^{l,p} \gets g(\mathbf{o}_i^{l,p},\mathbf{u}_i^{l,p}+ \boldsymbol{\epsilon})$}
        \EndFor
        \State{$\tau^l_p \gets \{ \mathbf{o}_{0}^{l,p}, \cdots, \mathbf{o}_{N}^{l,p}\}$}
    \EndFor
    \State{$\mathcal{O}_l \gets \{ \tau^l_1, \cdots, \tau^l_P\}$}
\EndFor

\State{$\phi \gets \phi(\{\mathcal{O}_1,\cdots,\mathcal{O}_L\})$} \Comment{Eq. (\ref{eq:augmeted running cost})}
\label{alg_line:terminal cost function}
\State{$\psi \gets \psi(\{\mathcal{O}_1,\cdots,\mathcal{O}_L\})$} \Comment{Eq. (\ref{eq:augmeted terminal cost})}
\label{alg_line:running cost function}
\label{alg_line:clustered_MPPI}
\State{$U\gets \text{Clust. MPPI}(f,\Sigma,\mathbf{x}_0,U_0,K,N,\lambda,\phi,\psi)$} \Comment{Alg. \ref{alg:Clustered MPPI}}
\end{algorithmic}
\end{algorithm}
For clarity and compactness, the DC-MPPI pseudo-code is only written for one dynamic obstacle.
The \textbf{for} loop starting in Line \ref{alg_lin:obs simulation} simulates how a dynamic obstacle move over time.
Lines \ref{alg_line:terminal cost function} and \ref{alg_line:running cost function} create a terminal and running cost function respectively based on the trajectories of the obstacles that were simulated.
The resulting terminal and running cost function only rely on the agent's state.
Finally, Line \ref{alg_lin:clustered_MPPI} sends all of the necessary data to Algorithm \ref{alg:Clustered MPPI} to calculate the final control input.

The improvements of both clustering the trajectories and accounting for dynamic obstacles can be seen in Figure \ref{fig:Great Example}. 
The reference control input is for the agent to go straight forward at 1 m/s. 
The deviation from the control input is only in the heading. 
The dynamic obstacle is directly between the agent and the goal causing a valley in the value function as seen in all of the bottom plots of Figure \ref{fig:Great Example}.
Only the baseline MPPI algorithm chooses a control input within this trough since it does not cluster.
The other algorithms are able to choose control inputs that produce safe trajectories if the obstacle was stationary.
However, the obstacle is dynamic.
Therefore, the region of bad trajectories expands in the DC-MPPI algorithm, but not for the Clustered algorithm.
This results in the DC-MPPI algorithm getting out of the way of the obstacle rather than going towards the goal like the clustered algorithm and the baseline.

\section{Experiments}\label{sec:Results}
All simulations were performed in MATLAB 2022b on a
laptop with Intel i7 processor and 32GB of RAM. 
MATLAB's implementation of DBSCAN was used for clustering. 
Furthermore, MATLAB's ode45 was used to rollout trajectories.

\subsection{Static Obstacles}
For the first experiment, an agent with Dubins car dynamics is maneuvering in a 2D plane and avoiding randomly placed obstacles.
The state of the agent is its position, $(x,y)$,and its direction of motion, $\theta$.
The dynamics of the system are given by

\begin{gather} \label{eq:1D Dynamics}
    \begin{bmatrix}
        \dot{x} \\ \dot{y} \\ \dot{\theta}
    \end{bmatrix}
    = \begin{bmatrix}
        v \cos(\theta) \\
        v \sin(\theta) \\
        0
    \end{bmatrix}
    + \begin{bmatrix}
        0 \\
        0 \\
        1
    \end{bmatrix} \omega.
\end{gather}
The linear velocity, $v$, is constant and positive, and the agent only has control over the angular velocity, $\omega$.
When choosing $\omega$, the agent must always satisfy $ 0 < R_{\min} \leq v / \omega$ where $R_{\min}$ is the minimum turning radius.

\begin{figure}
    \centering
    \includegraphics[trim={7cm 0cm 7cm 1cm},clip,width=.45\textwidth]{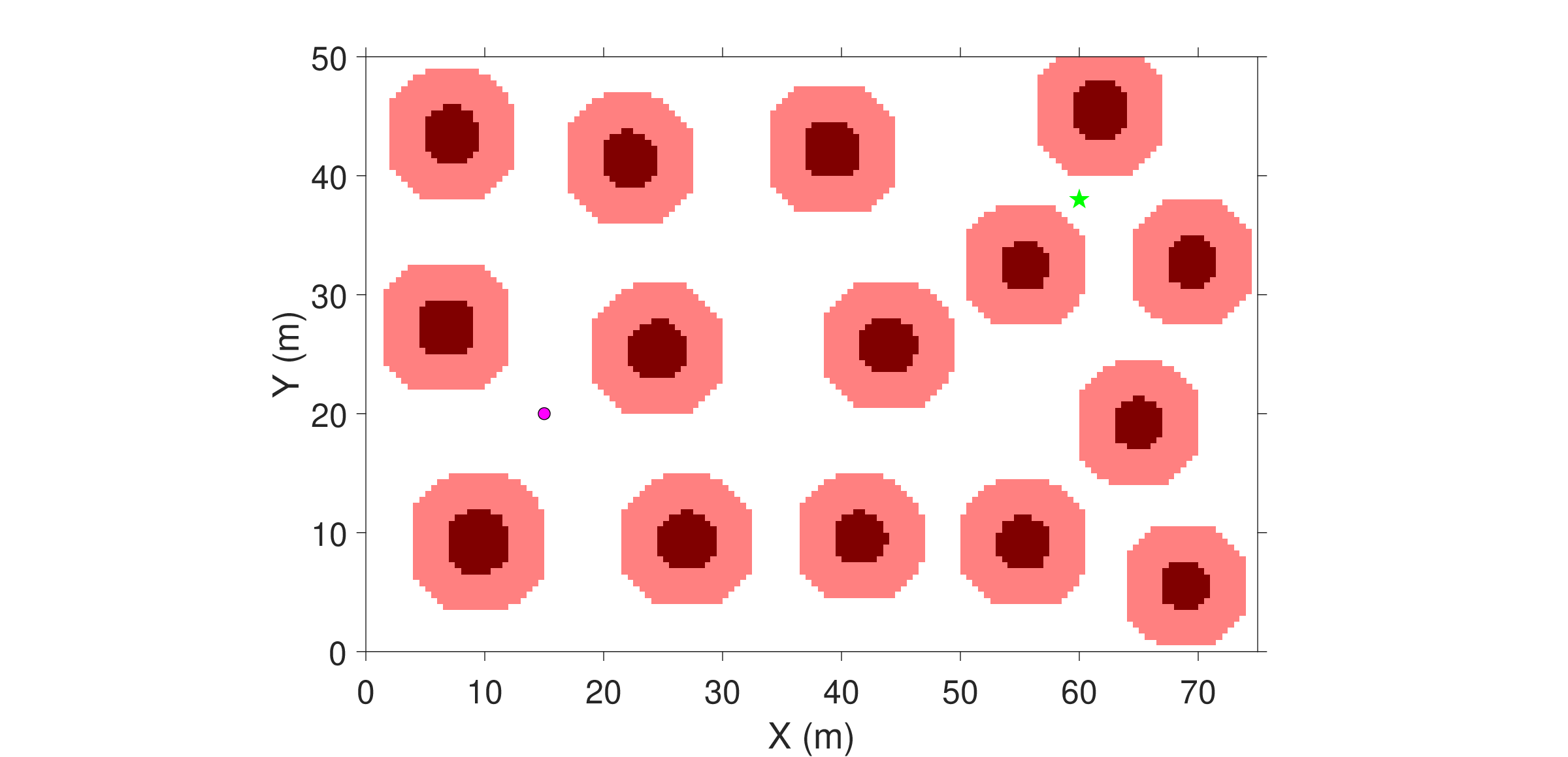}
    \caption{Random forest environment example. The agent is the magenta dot, the goal is the green star, and the obstacles are light and dark red.}
    \label{fig:Forest Example}
\end{figure}

The running cost and terminal cost are equal to the distance of, respectively, the current position and terminal position of the system from the goal position $(x_g,y_g)$ along with a penalty cost for being in collision with a static cost map: $\mathbf{1}(\mathbf{x})$. The collision cost is weighted by the hyper-parameter $\alpha > 0$.
\begin{gather}
    \phi(\mathbf{x}) = \psi(\mathbf{x}) = \sqrt{(x-x_g)^2+(y-y_g)^2}+\alpha\mathbf{1}(\mathbf{x})
\end{gather}

The proposed method was tested along with the standard MPPI formulation and another state-of-the-art method \cite{williams2018robust} using the authors' own implementation of each algorithm.
The start and goal positions were randomly generated for locations that were not in collision of any static obstacles within a randomly generated environment.
The obstacles are static and non-uniform in shape, and an example can be seen in Figure \ref{fig:Forest Example}.
The first iteration of MPPI is warm-started with a straightforward initial guess, but all subsequent MPPI calculations are initialized with the previous iteration's solution.
500 trajectories are simulated and the perturbation from the control reference is kept constant for the entire trajectory.
The hyper-parameters were set $\lambda=1$, $\alpha=1,000$, $\Sigma=0.1$. 
Lastly, the algorithms were tested with 3 different noise profiles: noiseless, input noise, process and input noise.

The algorithms were evaluated in 1000 different randomly generated environments and report the number of collision occurred, percentage failure which includes crashing and not reaching the goal in a timely manner.
The authors define a timely manner as being less than the triple the time it would take the agent to go around the entire map without any obstacles. 
The results can be seen in Table \ref{tab:Results 1D Static}.
For the static trials, DC-MPPI was not included since it produces the same output as Clustered MPPI when there are no moving obstacles.
It can be seen that the proposed method reduces the number of collisions and number of failures for all noise configurations while adding minimal computation time. 
\begin{figure*}
    \centering
    \includegraphics[trim={0.5cm 1cm 0cm 3cm},clip,width=.85\textwidth]{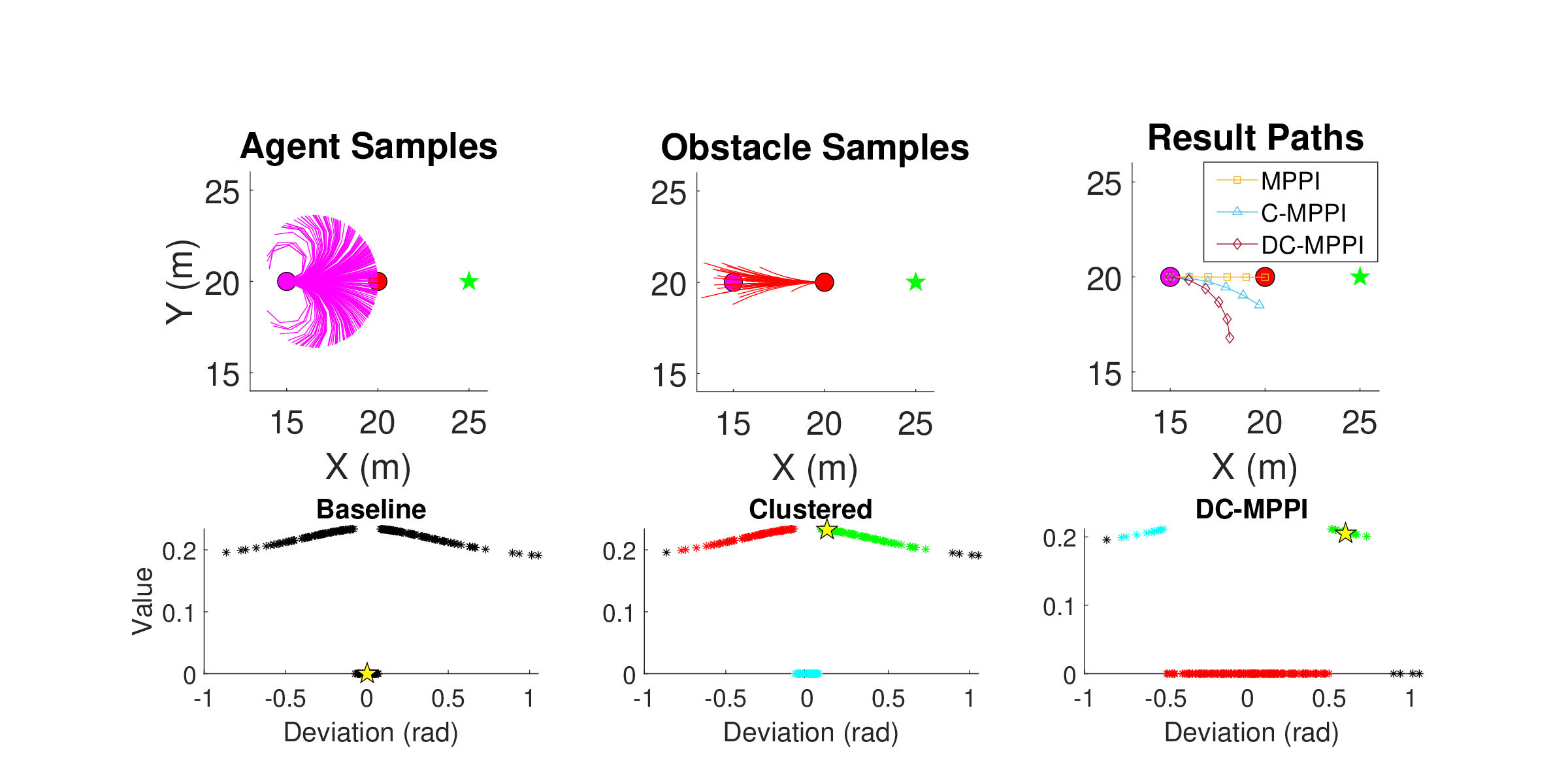}
    \caption{Dynamic obstacle example where the agent is a magenta circle, the obstacle is a red circle traveling in the negative $x$ direction, and the goal is the green star. 
    Top left are all the simulated trajectories of the agent. 
    Top middle are the simulated obstacle trajectories. 
    Top right are the resulting paths from each algorithm. 
    The bottom row are the plots of the value functions with respect to the deviation from the reference control input. 
    Different clusters of trajectories are denoted in different colors.
    The resulting control input deviation is denoted with a yellow star.}
    \label{fig:Great Example}
\end{figure*}
\begin{table}
    \centering
    \begin{tabular}{||c|c|c|c|c||}
        \hline
        Test & Algorithm & \# Collisions & \% Failure & Avg. \\
        & & & Time (ms) & Comp. \\
        \hline \hline
        \multirow{3}{*}{No Noise} & Baseline & 11 & 1.2 & 190 \\ \cline{2-5}
        & Tube-MPPI \cite{williams2018robust} & 9 & 1.2 & 364\\ \cline{2-5}
        & Clustered & 6 & 0.9 & 193 \\ \hline \hline
        \multirow{3}{*}{Control Noise} & Baseline & 59 & 6.0 & 195 \\ \cline{2-5}
        & Tube-MPPI \cite{williams2018robust} & 44 & 4.5 & 376\\ \cline{2-5}
        & Clustered & 26 & 2.9 & 199 \\ \hline \hline
        \multirow{2}{*}{Control and } & Baseline & 48 & 4.9 & 355 \\ \cline{2-5}
        & Tube-MPPI \cite{williams2018robust} & 49 & 5.0 & 376\\ \cline{2-5}
        Process Noise & Clustered & 37 & 3.9 & 360 \\ \hline
    \end{tabular}
    \caption{Quantitative performance over 1000 runs.}
    \label{tab:Results 1D Static}
\end{table}
\begin{figure}
    \centering
    \includegraphics[trim={7.1cm 0.25cm 7.5cm 1cm},clip,width=.45\textwidth]{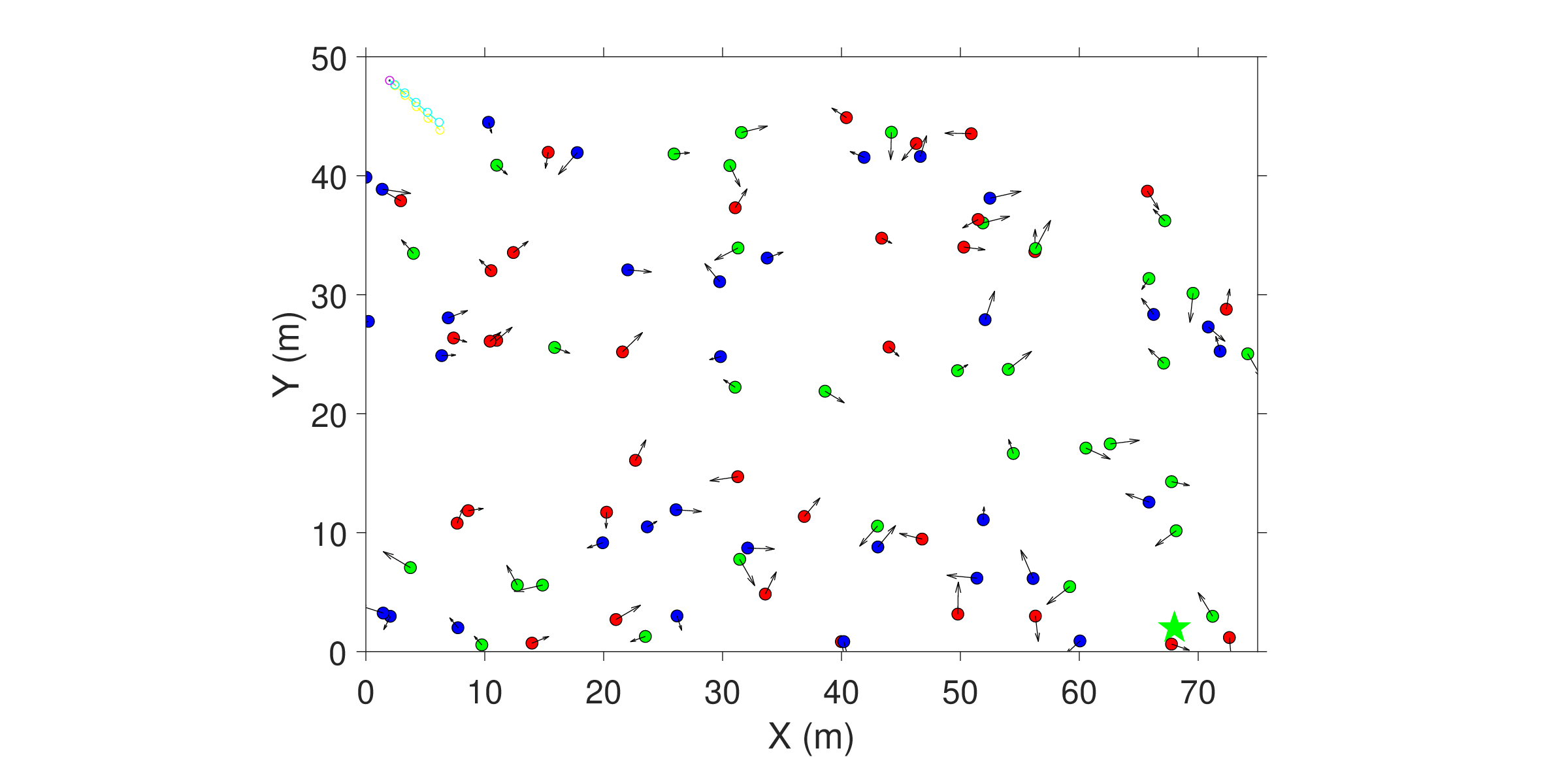}
    \caption{Dynamic environment example. Agent is magenta circle, obstacles are red, green and blue circles with their direction of motion indicated by black arrows. The goal is the green star. The yellow and cyan lines are the first iteration of DC-MPPI with the yellow being the reference input and the cyan being the result.}
    \label{fig:Dynamic Enviornment Example}
\end{figure}
\subsection{Dynamic Obstacles}
For the dynamic environment, the agent has the same dynamics as the previous experiment.
The state of the agent is its position $(x,y)$ and heading $\theta$ and respective velocities: $v_x,v_y,v_\theta$.
The control inputs are $u_x,u_y, u_\theta$.
The environment was populated with 100 dynamic obstacles.
These dynamic obstacles have Dubins car dynamics from Eq. (\ref{eq:1D Dynamics}) and they all have a 1 meter collision radius.
The starting locations of the obstacles are uniformly distributed over $[0, 75]\times [0,50] m^2$ space and the orientations are uniformly distributed over $[0,2 \pi)$ rad.
The linear and angular velocities are similarly distributed with $v \in [0.5, 1.5]$ m/s and $\omega \in [-0.5, 0.5]$ rad/s.
The obstacles' velocities are not explicitly known to the agent, but a Gaussian distribution is known of the potential linear and angular velocities of the obstacles is known.
This distribution is used to sample potential obstacle trajectories as seen in Figure \ref{fig:Great Example}.
The first time step of the environment can be seen in Figure \ref{fig:Dynamic Enviornment Example}.
Additionally, an example of how trajectories are sampled for the agent and obstacles and the samples effect on the value function of the separate algorithms can be seen in Figure \ref{fig:Great Example}.

The same hyper-parameters were used in this experiment as the previous one.
For DC-MPPI, 25 samples of each obstacle were used and $\beta=10$.
Using the baseline MPPI algorithm resulted in 3 instances of obstacle collisions whereas DC-MPPI had no collisions.
Without the dynamic obstacle sampling, the Clustered algorithm performed better than MPPI with 2 collisions.
It can be seen in the video\footnote{Video link: \url{https://youtu.be/G68DbQO1ouM}} that both the baseline and Clustered MPPI algorithms collide with moving obstacles when an obstacle passes through the reference path.
It should also be noted that Clustered MPPI has marginal impact on the computational time required with an average 288 milliseconds per time step for Clustered MPPI compared to 283 milliseconds per time step for the baseline algorithm.
This is in contrast to DC-MPPI which averaged 454 milliseconds.
The increase in computation time is clearly worth the decrease in collisions.
Additionally, the a-priori knowledge of the obstacles velocities does not diminish the applicability of this paper's novel algorithm to the real world.
When applied to a real system, a simple Extended Kalman Filter \cite{bar2004estimation} would be sufficient to generate a distribution to sample obstacle trajectories from.
\section{Conclusion}\label{sec:Conclusion}
In this paper, two improvements upon MPPI were presented. 
The first of which is clustering trajectories so that the resulting control input sequence is prevented from producing unsafe trajectories.
This improvement is specifically meant to separate clusters of good trajectories thereby preventing an unsafe average between neighboring clusterings.
It was shown that the clustering step adds little computation time to the MPPI algorithm, and demonstrated clear cases where the baseline fails.
The efficacy of DC-MPPI depends greatly on the hyper-parameter selection for DBSCAN.
For high-dimensional control input spaces, holes can appear in the middle of a cluster. 
Therefore, hyper-parameters should be selected where clusters stay relatively small ($< 20$\% of the trajectories) to prevent this from happening.

The second is creating a cost function that accounts for dynamic obstacles.
Adding dynamic obstacles allows for MPPI to prevent collisions where fast replanning is not sufficient.
In the experiments, it was shown that the proposed improvements reduced the number of collisions with a slight increase in the overall computation time.
With a GPU speedup, this additional time would significantly reduce.

Importantly, the proposed improvements can be added to the majority of MPPI variants like \cite{yin2022risk} without much retooling.
As seen in Algorithms \ref{alg:Clustered MPPI} and \ref{alg:DC-MPPI}, the novel algorithms can be considered a pre-processing step to the original MPPI algorithm.
\section*{APPENDIX}
\begin{proof}
Let one consider the distribution $\mathbb{Y}$ whose probability density function is given by Eq. (\ref{eq:Truncated Gaussian PDF}).
For $y(v)$ to be a valid PDF, the following must be true
\begin{align} 
    1 = \upsilon &\int_\Omega \exp \left ( -\frac{1}{\lambda} J(V)\right ) p(V) \mathrm{d}\Omega \notag \\ 
    \label{eq:constant constraint} &+ \sigma \int_{\Omega^c} \exp \left ( -\frac{1}{\lambda} J(V)\right ) p(V) \mathrm{d}\Omega^c.
\end{align}
Because of the constraint in Eq. (\ref{eq:constant constraint}) on $\upsilon$ and $\sigma$, the variables are dependent upon one another.
A new importance sampling scheme using $y(V)$ can be made through the following,
\begin{align}
    \mathbf{u}^*_i =& \int y(V) q(V) \mathbf{v}_i \mathrm{d}V \notag \\
    =& \int \frac{y(V)}{p(V)}\frac{p(V)}{q(V)} q(V) \mathbf{v}_i \mathrm{d}V \notag \\
    =& \upsilon \int_{\Omega} \frac{\exp \left ( -\frac{1}{\lambda } J(V) p(V) \right )}{p(V)}\frac{p(V)}{q(V)} q(V) \mathbf{v}_i \mathrm{d}\Omega \notag \\
    &+ \sigma \int_{\Omega^c} \frac{\exp \left ( -\frac{1}{\lambda } J(V) p(V) \right )}{p(V)}\frac{p(V)}{q(V)} q(V) \mathbf{v}_i \mathrm{d}\Omega^c \notag \\
    =& \upsilon \int_{\Omega} \exp \left ( -\frac{1}{\lambda } J(V) \right )\frac{p(V)}{q(V)} q(V) \mathbf{v}_i \mathrm{d}\Omega \notag \\
    \label{eq: another ratio}
    &+ \sigma \int_{\Omega^c} \exp \left ( -\frac{1}{\lambda } J(V) \right ) \frac{p(V)}{q(V)} q(V) \mathbf{v}_i \mathrm{d}\Omega^c
\end{align}
The ratio $p(V)/q(V)$ was defined in Eq. (\ref{eq:natural over good weighting}).
Therefore, a new weighting function $w^*(V)$ is used and defined as
\begin{align} \label{eq:weighting function proof}
    w^*(V) =& \begin{cases}
         \upsilon w(V), \quad V \in \Omega \\
         \sigma w(V), \quad V \in \Omega^c 
    \end{cases} \\
    w(V) =& \frac{1}{\rho} \exp \left ( -\frac{1}{\lambda} J(V) + Q(V) \right ) \\
    \rho =& \upsilon \int_\Omega \exp \left ( -\frac{1}{\lambda} J(V) + Q(V) \right ) \mathrm{d}\Omega \notag \\
    &+ \sigma \int_{\Omega^c} \exp \left ( -\frac{1}{\lambda} J(V) + Q(V) \right ) \mathrm{d}\Omega^c\\
    Q(V) =& \sum_{i=0}^{N-1} \frac{1}{2}\mathbf{u}^{(k)T}_i \Sigma^{-1} \mathbf{u}_i^{(k)} 
     - \mathbf{v}_i^{(k)T} \Sigma^{-1} \mathbf{v}^{(k)}_i.
\end{align}
As previously mentioned, there is only 1 degree of freedom for the variables $\upsilon$ and $\sigma$.
Thus the limit can be taken with respect to $\sigma$ approaching $0$.
Taking this limit with Eq. (\ref{eq: another ratio}) results in
\begin{align}
    \lim_{\sigma\to 0}(\mathbf{u}_i^*) = \upsilon \int_{\Omega} w(V) q(V) \mathbf{v}_i \mathrm{d}\Omega
\end{align}
To ensure $y(V)$ is still a valid PDF, Eq. (\ref{eq:constant constraint}) must still hold resulting in
\begin{align}
    \frac{1}{\upsilon} = &\int_\Omega \exp \left ( -\frac{1}{\lambda} J(V)\right ) p(V) \mathrm{d}\Omega \notag \\ 
    &+  \frac{\lim_{\sigma \to 0}(\sigma)}{\upsilon}\int_{\Omega^c} \exp \left ( -\frac{1}{\lambda} J(V)\right ) p(V) \mathrm{d}\Omega^c \notag \\
    = &\int_\Omega \exp \left ( -\frac{1}{\lambda} J(V)\right ) p(V) \mathrm{d}\Omega. \\
    \rho = &\upsilon \int_\Omega \exp \left ( -\frac{1}{\lambda} J(V) + Q(V) \right ) \mathrm{d}\Omega
\end{align}
With the normalizing constants derived, the final weighting for importance sampling within $\Omega$ becomes
\begin{align}
    w^*(V) &=
       \frac{1}{\eta} \exp \left ( -\frac{1}{\lambda} J(V) + Q(V) \right ) \\
       \eta &= \int_\Omega w^*(V) \mathrm{d} \Omega
\end{align}
and $0$ otherwise.
Thus it has been shown that $y(V)$ has the needed properties to be a valid PDF, and it is a PDF that is strictly positive when $p(V)$ is non-zero. 
With all conditions satisfied for $y(V)$ to be a valid PDF for importance sampling with respect to $p(V)$, the proof is complete.
\end{proof}
\bibliographystyle{IEEEtran}
\bibliography{sample}

\end{document}